 \newcommand{\y}{Y\-Ba$_{2}$\-Cu$_{3}$\-O$_{7-\delta}$}
\def\Bp{B_{\text{cp}}}

\def\Ba{B_a}

\newcommand{\f}[1]{Fig.~\ref{#1}}
\newcommand{\eq}[1]{Eq.~(\ref{#1})}

\def\s2{w/\sqrt{2}}
\def\d{\partial}
\def\be{\begin{equation}}
\def\ee{\end{equation}}
\def\bea{\begin{eqnarray}}
\def\eea{\end{eqnarray}}
\def\l({\left(}
\def\r){\right)}

\documentstyle[prl,aps,multicol]{revtex} 

  \renewcommand{\narrowtext}{\begin{multicols}{2} \global\columnwidth20.5pc}
  \renewcommand{\widetext}{\end{multicols} \global\columnwidth42.5pc}

\input{psfig}

\draft
\begin{document}
\title {Central peak position in magnetization loops of high-$T_c$
superconductors}  
\author{D.~V. Shantsev$^{1,2}$,
M.~R.~Koblischka,$^{1,}$\cite{0} Y.~M.~Galperin$^{1,2}$, T.~H.~Johansen$^1$,
P. Nalevka$^3$, and M. Jirsa$^3$}

\address{$^1$Department of Physics, University of Oslo, P. O. Box 1048 
Blindern, 0316 Oslo, Norway\\
$^2$
A. F. Ioffe Physico-Technical Institute, Polytechnicheskaya 26, 
St.Petersburg 194021, Russia\\
$^3$ ASCR, Institute of Physics, Na Slovance 2, CZ-18040 Praha, Czech Republic}

\date{\today}

\maketitle

\begin{abstract}
Exact analytical results are obtained for the 
magnetization of a superconducting thin strip with a general behavior
$J_{c}(B)$ of the critical current density.    
We show that within the critical-state model the magnetization as
function of applied field, $\Ba$, has an extremum located exactly at
$\Ba=0$. 
This result is in excellent agreement with presented
experimental data for a \y\ thin film. 
After introducing granularity by patterning the film, the central peak
becomes shifted to positive fields,  $\Bp>0$, 
on the descending field branch of the loop. 
Our results show that a positive  $\Bp$ is a definite signature of
granularity in superconductors. 

\end{abstract}

\pacs{PACS numbers: 74.25.Ha, 74.76.Bz, 74.80.Bj}
\narrowtext

%\section{Introduction}

The investigation of magnetic hysteresis loops (MHLs) is a widely used tool to
characterize superconducting samples, and in particular, to estimate the
critical current density and its dependence on magnetic field. An
ever-present feature of the MHLs is a  
peak in the magnetization located at an applied field $\Bp$ near
zero. The central peak is formed due to a field dependence of the 
critical current density, $J_{c}(B)$, monotonously decreasing at small
fields. When the sample is a long cylindrical body placed in a
parallel applied field, the peak position can be calculated analytically
within the critical-state model for several $J_{c}(B)$ 
dependences~\cite{JohBra}. One always finds on the descending field
branch that $\Bp<0$, and similarly $\Bp>0$ on the
ascending branch of large-field loops. In simple terms the shift is a
consequence of the local flux density, $B$, lagging behind the applied field.

Also experimentally there are abundant observations of negative $\Bp$
on the descending field branch. However, in some cases one finds $\Bp$
\ very close to zero or even shifted to the positive side 
so that the peak occurs before the remanent state is 
reached~\cite{McHenry,Cim,Mul97,MK-APL,MK-JAP}. The explanations
for this behavior are controversial. On one hand, a decreasing sample
thickness is known from experiment to shift the peak toward
$\Bp=0$~\cite{Daumling,Daum}. This is also 
in agreement with recent numerical 
results obtained for the critical-state model with $B$-dependent 
$J_{c}$~\cite{Daum,McDonald}. However, the numerical calculations have
not predicted 
sufficiently large shifts to bring the peak to the anomalous side of the loop.

A different explanation of the shift in $\Bp$ was suggested for
granular materials,~\cite{Evetts} where a demagnetization effect of the
grains can be important. The magnetization of a granular superconductor is
determined by both intra- and inter-grain currents. The latter represent
large-size current loops, and can give a major contribution to the magnetic
moment. These inter-grain currents are essentially determined by the
magnetic field just at the grain boundaries. Those local fields are, in turn,
strongly influenced by the intra-grain currents, and can vary {\em
ahead} of the local $B$ expected without granularity. As a result, the 
central peak is encountered earlier in the MHL. Models allowing
quantitative MHL 
calculations with account of the grain-induced demagnetization effect~\cite
{Mul97,Dyach96} successfully reproduced the peak at $\Bp>0$.

Although the significant influence of both sample shape and
sample granularity on the MHL is generally recognized, the interplay
between them has never been addressed. In particular, it is not known if a
positive $\Bp$ is accessible for thin enough samples, or if $\Bp>0$ on
the descending field branch is a definite signature of sample 
granularity. The present work aims to answer this question.
We show first analytically that for {\em any} $J_c(B)$ the central
peak is located exactly at $\Bp=0$ for a thin uniform strip, a result
we also confirm by experiment. The key role of granularity is then
demonstrated by measurements on an artificially granular thin film,
where we find $\Bp>0$.  
\begin{figure}[h]
\centerline{ \psfig{figure=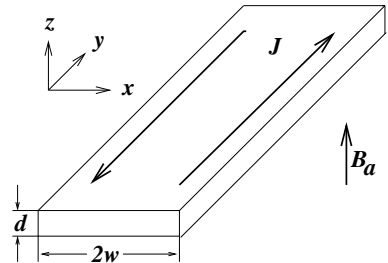,width=5cm}} 
\caption{Superconducting strip in applied magnetic field. }
\label{strip}
\end{figure}
Consider a long thin superconducting strip with edges located at $x=\pm w$,
the $y$-axis pointing along the strip, and the $z$-axis normal to the strip
plane, see \f{strip}. The magnetic field,  $B_{a}$, is applied along
the $z$-axis, so screening currents flow in the
$y$-direction. Throughout the paper  
$B$ is the $z$-component of magnetic induction in the strip plane. The sheet
current is defined as $J(x)=\int {j(x,z)\, dz}$, where $j(x,z)$ is the current
density and the integration is performed over the strip thickness, $d\ll w$.

{}From the Biot-Savart law for the strip geometry, the flux density is
given by~\cite{BrIn} 
\begin{equation}
B(x)-B_{a}=-\frac{\mu _{0}}{\pi }\int_{-w}^{w}\frac{J(u)\,du}{x-u}\, .
\label{B}
\end{equation} 
Assume that the strip is in a fully penetrated state, i.e., the
current density is everywhere equal to the critical one, 
$J(x)={\rm sign}(x)\,J_{c}[B(x)]$. This state can be reached after
applying a very large 
field, and then reducing it to some much smaller value. The field
distribution then satisfies the following integral equation,
\begin{equation}
B(x)-B_{a}=-\frac{\mu _{0}}{\pi }\int_{0}^{w}
  \frac{J_{c}[B(u)]}{x^{2}-u^{2}}\,udu\, .
\label{ie}
\end{equation}

In the remanent state, $B_{a}=0$, the flux density profile $B(x)$ has an
interesting symmetry. This is seen by changing the integration variable 
in Eq.~(\ref{ie}) from $u$ to $v=\sqrt{w^{2}-u^{2}}$. We then obtain 
\begin{equation}
B(x)-B_{a}=\frac{\mu _{0}}{\pi
}\int_{0}^{w}\frac{J_{c}[B(\sqrt{w^{2}-v^{2}})]}{w^{2}-x^{2}-v^{2}}
\,v \, dv \,.  \label{tm3} 
\end{equation}
Substituting $x \rightarrow \sqrt{w^2-x^2}$ into \eq{ie} one obtains
a similar equation for $B(\sqrt{w^{2}-x^{2}})$,
\begin{equation}
B(\sqrt{w^{2}-x^{2}})-B_{a}=-\frac{\mu _{0}}{\pi }\int_{0}^{w}
\frac{J_{c}[B(u)]}{w^{2}-x^{2}-u^{2}}\,udu \,.  
\label{tm4}
\end{equation}
By comparing equations (\ref{tm3}) and~(\ref{tm4}) at $B_{a}=0$  we
conclude that 
\begin{equation}
B(x)=-B(\sqrt{w^{2}-x^{2}})\,
\label{tm5}
\end{equation}
is generally valid if $J_{c}$ depends only on the absolute value of
the magnetic induction,  
$J_{c}(|B|)$.
This symmetry is immediately evident for the case $J_{c}=$ const., i.e., the
Bean model, when Eq.~(\ref{ie}) reduces to 
\begin{equation}
B(x)=B_{a}+B_{c}\,\ln \frac{\sqrt{w^{2}-x^{2}}}{x},\quad B_{c}=\frac{\mu
_{0}J_{c}}{\pi }\,.  \label{Bean}
\end{equation}
In a similar way one can prove also another symmetry relation valid at 
$B_{a}=0$, namely
\begin{equation}
{\cal D}(x)={\cal D}(\sqrt{w^{2}-x^{2}})\,,  \label{tm6}
\end{equation}
where ${\cal D}(x)=\partial B(x)/\partial B_{a}$.

\begin{figure}[h]
\centerline{ \psfig{figure=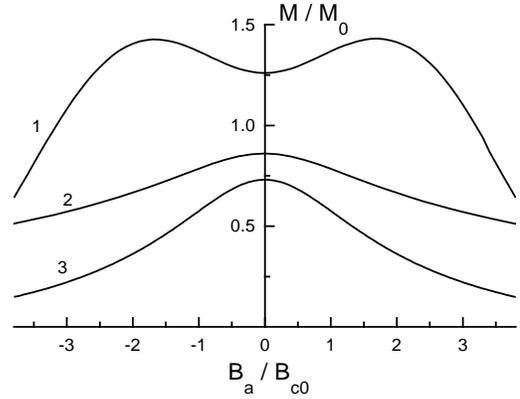,width=8cm}} 
\caption{Central part of descending branches of the MHL for different
$J_c(B)$ dependences, $J_c/J_{c0}={\cal F}(|B|/B_{c0})$. 
 (1)~-- ${\cal
F}(b)=(1+2.5b^2)e^{-b^2/4}$; (2) -- ${\cal F}(b) = (1 + 0.25 b)^{-1}$;
(3) --  ${\cal F}(b)= e^{-b/2}$. Here    
 $B_{c0} = \mu_0 J_{c 0}/\pi$, $M_0 = w^2 J_{c0}$. }
\label{max}
\end{figure}

Consider now the magnetic moment per unit length of the
strip,   
$M = 2\int_{0}^{w} J(x) x\, dx $.
Differentiating $M$ with respect to $\Ba$ and taking into account that
$B(x)$ changes sign at $x=w^*=\s2$ one has after splitting 
the integral into two parts 
\begin{equation} %\label{}
\frac{\d M}{\d \Ba} 
=2\left(\int_0^{w^*}-\int^w_{w^*} \right) \frac{\d J_c(|B(x)|)}{\d
|B|}{\cal D} (x)  \, x \,dx \, .
\end{equation}
Then, replacing in the second
integral $x$ by $\sqrt{w^2-x^2}$ and using Eq.~(\ref{tm5}), we come to
\be
 \frac{\d M}{\d \Ba} = 0\, \quad  {\rm at} \quad \Ba=0 \ .
\ee
Consequently, on a major MHL the magnetic moment has an extremum in
the remanent state  for {\em any} $J_{c}(B)$-dependence. This is 
illustrated in Fig.~\ref{max} showing the central part of MHLs obtained by 
an iterative numerical
solution of Eq.~(\ref{ie}) for three $J_{c}(B)$ dependences. The $M(B_{a})$
extremum depends on the type of $J_{c}(B)$ dependence. It is a maximum
for a decreasing $J_{c}(B)$, and a minimum if $J_{c}(B)$ has a 
pronounced second peak, the so-called fishtail behavior.

Two \y\ epitaxial thin films were chosen for a comparative study; 
(A) a uniform film of regular shape, and (B) a film with artificial
granularity. Sample (A) was 
prepared by laser ablation on an MgO substrate~\cite{Shen}. The thickness was
200 nm, and the sample was patterned by chemical etching into a rectangular
shape of dimensions 0.66 $\times $ 1.4 mm$^{2}$. The homogeneity of the
sample was tested by magneto-optical imaging, where it showed complete
absence of any visible defects~\cite{JPC}. 

\begin{figure}[h]
\centerline{\psfig{figure=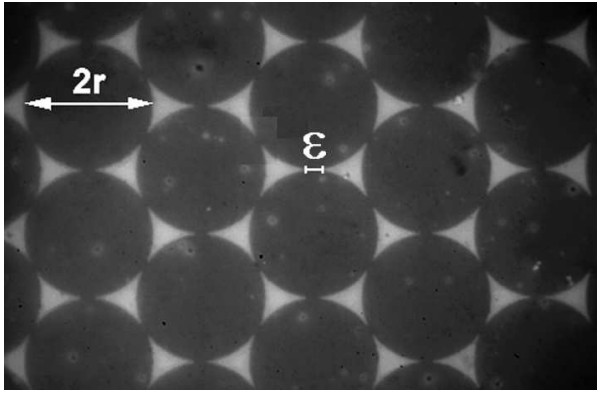,width=8cm}}  
\caption{Polarization image of the
sample (B) with artificial granularity. The black disks are
superconducting and have a diameter of $2r=50\ \mu$m. The width 
of the contact region between the disks is $\epsilon=3.5\ \mu$m. }
\label{pattern}
\end{figure}

Sample (B) was laser ablated to a 150~nm thickness on a LaAlO$_{3}$
substrate. The film was
patterned  by means of electron beam lithography into a hexagonal
close-packed lattice of disks with $2r=50~\mu$m diameter. The disks are
touching each other at the circumferences in order to enable the flow of
inter-disk currents. The width of the contact region is
$\epsilon=3.5~\mu$m, see Fig.~\ref{pattern}. The superconducting disks
can be considered as 
grains and the connections between them as inter-grain junctions. The
overall size of the sample was 4$\times $4~mm$^{2}$, comprising $\approx 8000
$~disks. The transition temperature after the patterning process is 
83~K.

The magnetization measurements were performed using a
vibrating sample magnetometer (VSM) PAR Model 155 with computerized control and
data processing. The magnetic field is generated by a conventional magnet
with $B_{{\rm a,max}}$ $\pm $2 T. The field was applied parallel to
the $c$ axis, which is  
perpendicular to the film plane. 
The magnetization loops were always measured after zero-field cooling of the
sample. The maximum applied field for all measurements was significantly
higher than the field giving a fully penetrated state, as verified by the
magneto-optical method. 
To focus on our main issue, only the data
measured on the descending field branch of the MHLs near $\Ba=0$ are presented.
\begin{figure}[h]
\centerline{ \psfig{figure=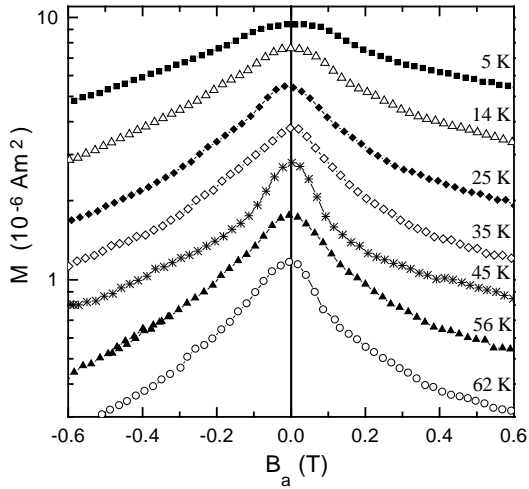,width=8cm}} 
\caption{Central part of descending branches of the MHL for a thin
YBCO strip for 
different temperatures. The peak in magnetization is located always at
zero applied field. }
\label{F4}
\end{figure}

Figure\ \ref{F4} shows the central part of the MHLs for the uniform
film (A). The measurements were carried out over a wide range of
temperatures in order to probe different $J_c(B)$ dependences. There
is evidently here a pronounced central peak at all temperatures. Within
the experimental resolution the position of the peak 
is located at zero applied field. The
observation that the peak remains at $\Ba=0$ over the entire
temperature range is in full agreement with our general analytical
result. 

Note that the result was derived for an infinite
strip fully penetrated by the magnetic field. In a thin sample, in contrast
to long cylinders, the flux fronts move in response to a changing
applied field at an exponential rate~\cite{BrIn}. Therefore, a fully
remagnetized state 
is reached quickly after reversing the direction of a field sweep, and our
experimental conditions are consistent with the assumptions made in the
theory. The only exception is the finite length of the strip. However,
our results show that it is not a crucial factor for the peak
position.  
\begin{figure}[h]
\centerline{\psfig{figure=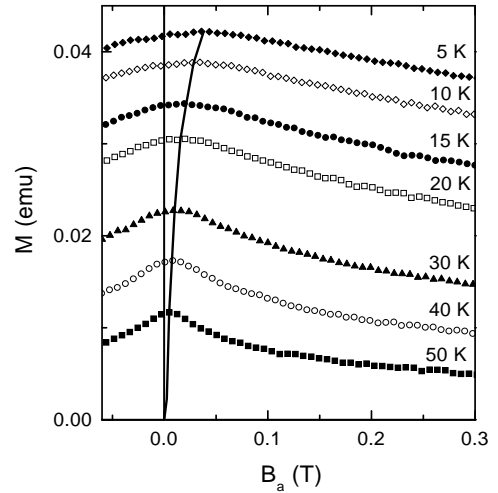,width=8cm}} 
\caption{Central part of the descending branch of MHLs for 
the sample (B) with artifical granularity. The solid line indicates
position of the magnetization peak 
which is located at positive fields at all the temperatures.}
\label{disks}
\end{figure}

The MHLs for the sample (B) with artificial granularity are shown in
Fig.~\ref{disks}. The central peak displays here different behaviour
as compared to the uniform film (A). For all   
temperatures $\Bp>0$ on the descending 
field branch. The shift of the central peak becomes more 
and more pronounced with decreasing temperature reaching 
$\Bp=+40$~mT at $T=5$~K.

In this sample,
the ``grains'' and their interconnections are of the same material, and
hence have equal critical current density.
Due to the patterning, the spatially
averaged inter-grain current density $J_{c}^{tr}$ is less than the current
density, $J_{c}^{gr}$, of the superconducting material itself. 
For the hexagonal pattern, see Fig.~\ref{pattern}, one has
$J_{c}^{tr}=\eta J_{c}^{gr}$, where $\eta$ depends on the direction of
$J_{c}^{tr}$, within the limits $\epsilon /2r<\eta <\epsilon
/\sqrt{3}r$. 
Futhermore, the magnetic moment $M^{gr}$ due 
to intra-grain currents and moment $M^{tr}$ due to inter-grain
currents are related as 
$$M^{gr}/M^{tr} = (J_{c}^{gr}/J_{c}^{tr})(r/R) \approx 2r^{2}/(\epsilon R), $$
where $R$ is the overall size of the sample.
In the present case this
ratio is 1/6, showing that the inter-grain contribution to the total
magnetic moment is dominant.
A considerable grain-induced demagnetization effect is therefore expected for this sample. Its granularity, being the only essential difference from the 
uniform film, is the only possible origin of the observed shift to $\Bp>0$. 

The superconducting Bi-2223 tapes are known to have a granular
microstructure and the two contributions to the magnetic moment are of
comparable magnitude~\cite{Mul97}.  
The grain demagnetization effect has been suggested as an explanation
for the observed shift of the peak to $\Bp>0$ on the descending MHL
branch in these materials~\cite{Cim,Mul97,MK-APL,MK-JAP}. Our present
results give strong support to this physical picture. 
\begin{figure}[h]
\centerline{ \psfig{figure=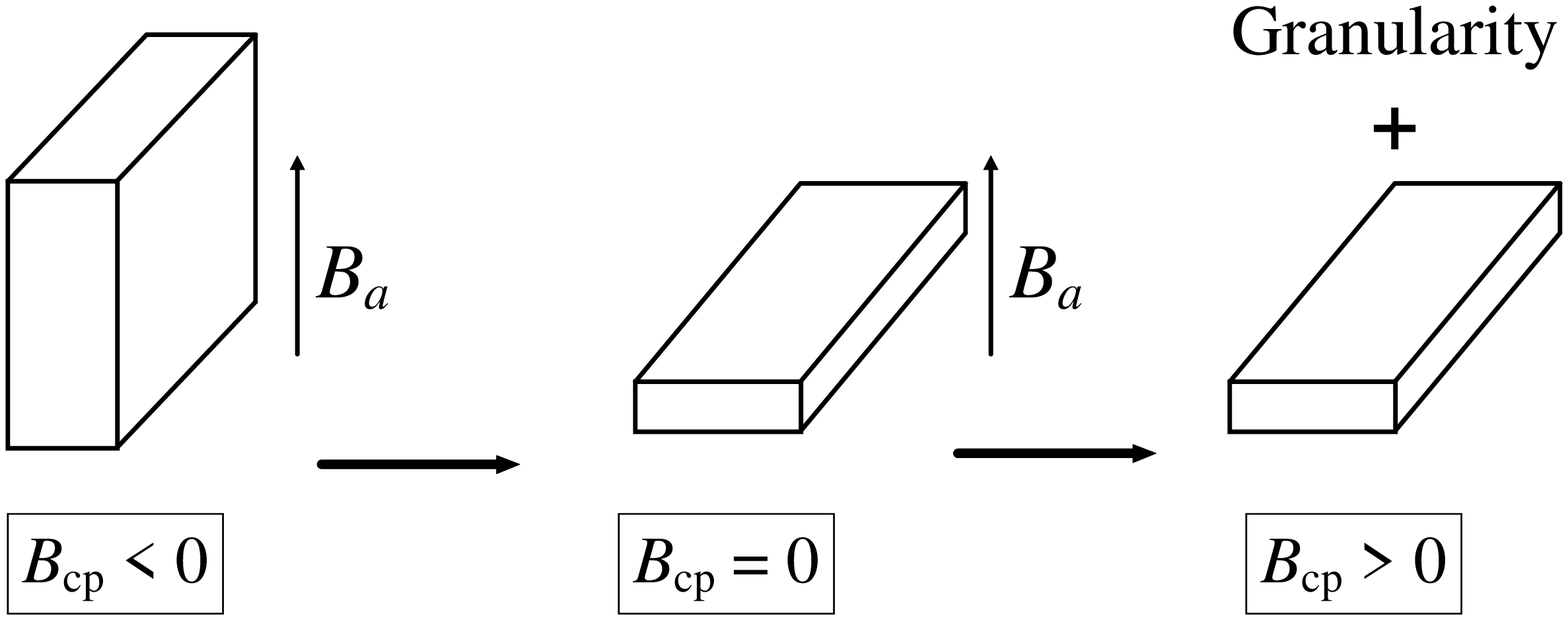,width=8.6cm}} 
\caption{A drawing illustrating the main conclusion of the paper. Long
samples in parallel magnetic field show the peak on the descending field
branch of MHL at a negative field. As the sample height decreases the peak
is shifted towards zero. It is exact zero in the limiting case of a uniform
thin strip. A granular thin strip has the peak at positive fields. }
\label{final}
\end{figure}

In conclusion, we arrive at a following general scenario concerning the
central peak position, see Fig.~\ref{final}.
Long samples in parallel magnetic field show in
their MHLs a central peak at a negative applied field, i.e. after passing
through the remanent state on the descending field branch. As the sample
thickness decreases the peak position is shifted towards zero. It is
located exactly 
at $\Bp=0$, in the limiting case of a uniform strip of
infinitessimal thickness. Granularity always leads to a shift of
$\Bp$ in the positive direction on the descending field
branch. Thus, for a 
granular thin strip the peak is located at a positive field, i.e., before
the remanent state is reached. The origin of this effect is that
granularity induces demagnetization fields which strongly
modify the inter-granular currents via their $B$-dependence. 

%\acknowledgements  

We thank Y.\ Shen (NKT Research Centre, Br{\o}ndby,
Denmark) for the excellent uniform YBCO thin film as well as
B. Nilsson and T. Claeson (Chalmers, G{\o}teborg, Sweden) for the film
with artificial granularity. We 
acknowledge valuable discussions with Prof. M. Murakami (SRL/ISTEC, Div. 3, 
Tokyo, Japan) and L.\ P\accent23ust (Wayne State University, Detroit,
USA). The financial support from the Research Council of Norway, from 
the Russian National Program for Superconductivity, and from the grant No.
A1010512 of GA ASCR is gratefully acknowledged.

%\narrowtext

\widetext
\end{document}